\def\year{2019}\relax
\documentclass[letterpaper]{article} 
\usepackage{aaai19}  
\usepackage{times}  
\usepackage{helvet}  
\usepackage{courier}  
\usepackage{url}  
\usepackage{graphicx}  
\usepackage{tikz}
\usepackage{caption}
\usepackage{subfig}
\usepackage{comment}
\usepackage{listings}
\usepackage[ruled,linesnumbered,titlenumbered]{algorithm2e}
\usepackage{pgflibraryshapes}
\usepackage[inline]{enumitem}
\usepackage{alltt}
\usepackage{comment}
\usepackage{mdframed}

\newcommand{\citet}[1]{\citeauthor{#1}~(\citeyear{#1})} 
\nocopyright
 
\begin{document}
%
\title{Schemaless Queries over Document Tables with Dependencies}
\author{
Mustafa Canim\\
IBM Research
\And
Cristina Cornelio \\
IBM Research
\And 
Arun Iyengar\\
IBM Research
\And
Ryan Musa \\
Google, Inc.\thanks{work conducted while at IBM Research}
\And
Mariano Rodrigez Muro\\
Google, Inc.$^*$
}
\maketitle
\begin{abstract}
Unstructured enterprise data such as reports, manuals and guidelines often contain tables. The traditional way of integrating data from these tables is through a two-step process of table detection/extraction and mapping the table layouts to an appropriate schema. This can be an expensive process. 
In this paper we show that by using semantic technologies (RDF/SPARQL and database dependencies) paired with a simple but powerful way to transform tables with non-relational layouts, it is possible to offer query answering services over these tables with minimal manual work or domain-specific mappings. 
Our method enables users to exploit data in tables embedded in documents with little effort, not only for simple retrieval queries, but also for structured queries that require joining multiple interrelated tables.
\end{abstract}

\section{Introduction}

Enterprise data includes large volumes of unstructured documents in the form of reports, manuals, guidelines, etc. Tables are of major importance within these documents. Often, these tables contain reference data that are required by enterprise personnel (e.g., help-desk operators, technicians in assembly lines, etc.) and difficult to access and query given that they are buried within documents. 

The traditional way of extracting data from these tables is through table extraction and mapping~\cite{Liu:2007:TAT,Liu:2006:AET,Lipinski:2013,Pinto:2003:TEU,Tengli:2004:LTE,Ramel:2003:DER,Cafarella:2009:WES,Wang12}, that is, components dedicated to detect the presence of tables in documents, extract them in a structured format, and use methods for mapping tables and table content into global schemas~\cite{DBLP:conf/semweb/HassanzadehWRS15}. While there has been success in domain-specific methods to create these mappings in several areas~\cite{Burdick11,BalakrishnanCHHKLPPPRSSTVY10,BurdickEKLPRW14,XuBR16}, general techniques are not available. Today, it is common that people interested in data within tables have to resort to manual mapping techniques, an expensive and intractable process. 

In this paper we take a different approach to the problem of accessing data in these tables. Here we show that by using semantic technologies (RDF/SPARQL and database dependencies) paired with a simple but powerful way to transform tables with non-relational layouts, it is possible to do query answering over these tables with minimal manual work, domain adaptation/mapping, or even full knowledge of the arrangement of the tables by the user formulating the query. 
In particular we focus on types of queries in which the input data is not sufficient to define a single answer and/or involves  the  contents  of  multiple  tables. This last case is especially difficult since is not possible to provide an answer simply joining tables columns because to complete the keys it is necessary to recover values that are not directly specified in the input.

Our key contributions are as follows: \begin{enumerate*}[label=\textit{(\alph*)}]
\item We describe a classification based method to transform tables with complex layouts into ``flat'' layouts that are closer to a relational form (e.g., SQL-like) and allow the use of relational theory to discover structural meta-data about the tables (e.g. keys and inclusion dependencies) to enable complex query answering.
\item We propose a method to execute conjunctive queries on collections of tables with unknown schema. In particular, we use an RDF graph implementing a \emph{universal schema} to store and index all rows for all tables and enable structured queries using search and features of SPARQL. 
\item We use relational dependencies to understand when a query requires JOIN-ing multiple tables and use a query-rewriting style approach to execute such queries (that allows to find the missing components of the input). 
\end{enumerate*}

Our method enables users to exploit data in tables with minimal effort, not only for simple retrieval queries, but also for complex structured queries. The methods described here have been tested and validated in the context of collaboration with a large industrial partner. 


 For the rest of the paper we proceed as follows: In Section~\ref{sec:preliminaries} we introduce some background definitions; in Section~\ref{sec:system} we briefly summarize all the steps of our method; in Section~\ref{sec:flattening_familyDetection} we describe our approach to transform tables with non-relational layouts to relational-like layouts; in Section~\ref{sec:aboxes_description} we describe the RDF schema we use to store the data and meta-data for all tables in the corpus in a triple store; in Section~\ref{sec:queryanswering} we introduce our query answering approach; in Section~\ref{sec:evaluation} we describe the application of these techniques in a project with an industrial partner; finally, in Section~\ref{sec:conclusions} we describe how the techniques presented here could be further extended to support more complex queries or stronger semantic integration/understanding of tables.

\section{Preliminaries}\label{sec:preliminaries}

Now we present a few definitions that we use throughout the paper, in particular, the notions of \emph{relations} and \emph{dependencies}.

Intuitively, a relation corresponds to a ``flat table'' or an SQL table, with a list of column names and rows (formally \emph{tuples}). Each row in a relation holds the values for all the attributes of an object of interest, e.g., something concrete like a screw and its features (e.g., length, weight, identification code) or something abstract such as the time and place for an event. Formally, we define a \emph{relation} $R$ \cite{AbHV95} as an ordered list of \textit{attributes} $U$. A \textit{tuple} (i.e. row) over $R$ is a mapping from each attribute $U_i \in U$ to a value. An \emph{instance} of a relation $R$, is a set of tuples over $U$. The \textit{projection} of a relation $R$ over a subset of attributes $U'$ is written as $\pi_{U'}(R)$.

Intuitively, a {\it dependency} is a logical implication between columns of one or more relations (flat tables). More formally, dependencies describe the ways in which data is arranged in relations. Dependencies are used for many purposes, one of the most common ones is consistency checking of tables in databases. The two types of dependencies that we use in this work are \emph{functional dependencies} and \emph{inclusion dependencies}. 

A {\it functional dependency} from a set of columns $U'$ in a relation $R$ to a second set of columns $V'$ in a relation $S$ (written as $R:U' \rightarrow S:V'$ ) means that for any set of unique values for $U'$, there is a unique set of values for $V'$, i.e., $U'$ functionally determines $V'$. We might omit writing $R$ and $S$ when they are unambiguous from the context. One of the most common forms of functional dependencies are \emph{keys}; a key over a relation $R$ is a functional dependency $U' \rightarrow U$; 
 the attributes in $U'\subseteq U$ uniquely determines the rest of the attributes of $R$. For simplicity we may write $\mathrm{Key}(R, U')$. When $U'$ is a single attribute, we simply call it  a \textit{key}; when $U'$ comprises 2 or more attributes, we call it a \emph{composite key}.
Instead, an \textit{inclusion dependency} is an implication of the form $R:U' \subseteq S:V'$, indicating that the values for attributes $U'$ in relation $R$ are a subset or equal to the values for the attributes $V'$ over the relation $S$. Intuitively, this indicates that the attributes $U'$ and $V'$ share data and hence, are ``joinable''. A special case for inclusion dependencies are \emph{foreign keys}, where $V'$ is also a key for $S$.

\section{System Overview}\label{sec:system}
\begin{figure*}[!hbt]
    \centering
    \includegraphics[width=\textwidth]{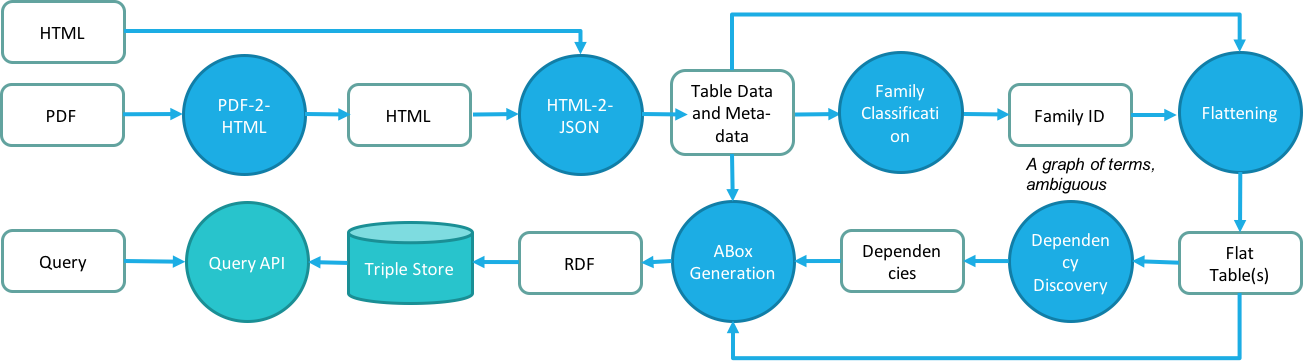}
    \caption{Ingestion and query answering system flow}
    \label{fig:ingestion}
\end{figure*}

The system is deployed as two services, an \emph{ingestion} service and a \emph{schemaless query API}, both depicted in  Figure~\ref{fig:ingestion}.
The system's ingestion flow starts by transforming PDF documents into HTML. We use IBM's document conversion services (under \emph{Watson Discovery}\footnote{\url{<https://www.ibm.com/watson/services/discovery/>}}) for this purpose. From the HTML we extract the tables and generate a JSON representation in which we do some minor normalization of the data (i.e., remove spanning cells). The features of the tables such as headers are first extracted and then tables are labeled with a \emph{table family} using a supervised classifier. The tables are then \emph{flattened} w.r.t. the family's layout and \emph{dependencies} are computed for the flat table. Finally, the flat table is stored as an ABox using a \emph{universal schema} approach, together with its corresponding dependencies and some additional meta-data (e.g., captions if available, header names, etc.). The RDF data is stored in a triple-store with support for search over triples (i.e., using search indexes). Last, the \emph{schemaless query API} is deployed over the triple store.

After this overview, we now proceed to describe in detail the main parts of the system.

\section{Compact Tables and Table Expansion}\label{sec:flattening_familyDetection}

Tables within documents can be arbitrarily complex due to the free nature of publication layouts. The work we present in this section aims at taming some of this complexity, particularly that arising from \emph{compact} table layouts. 

\begin{figure}[!htb]
\centering
\resizebox{\linewidth}{!}{
    \includegraphics{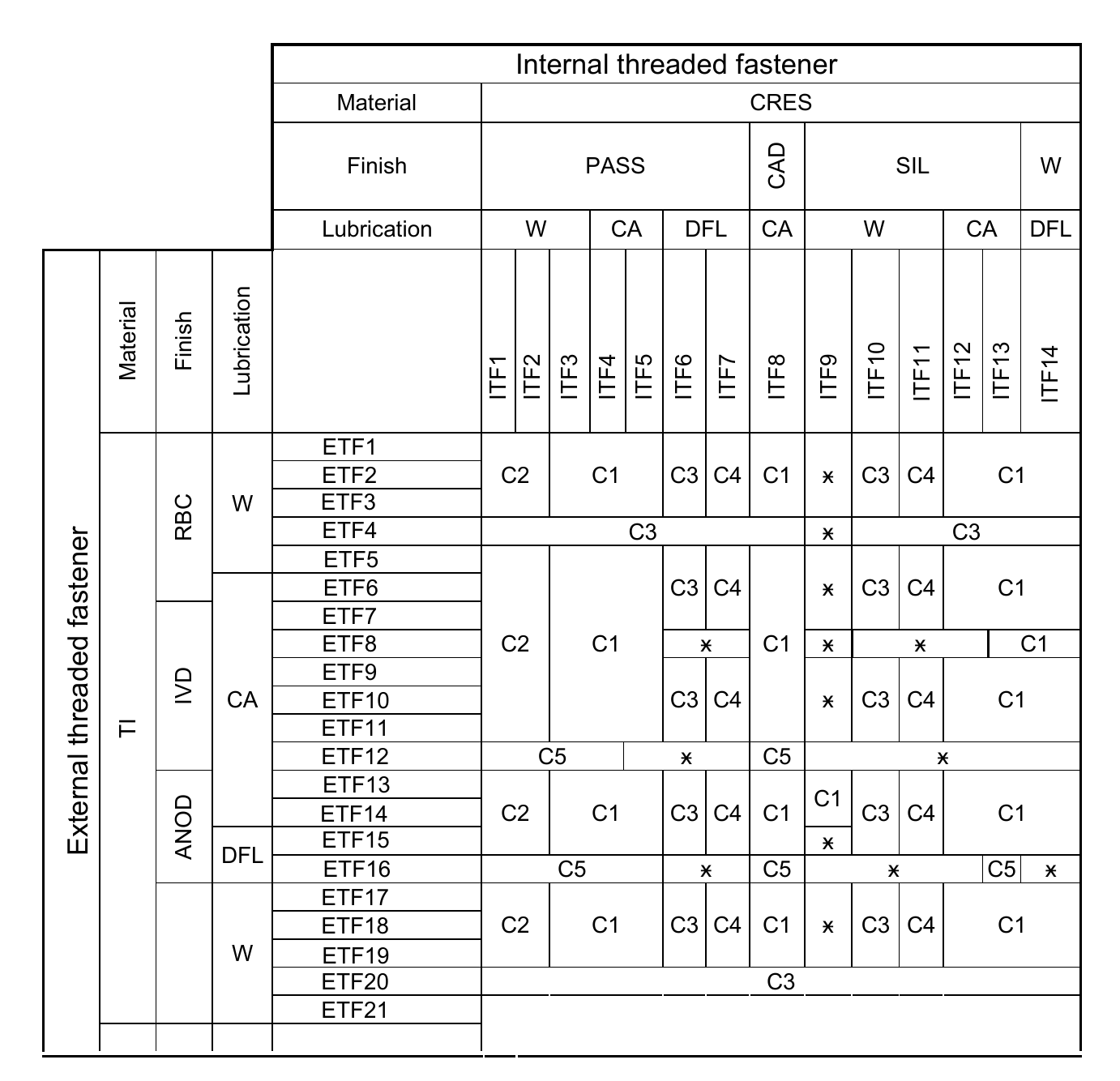}
    }
\caption{Compact table layouts that group values around keys located vertically and horizontally: Combination codes for Nut and Bolt pairs.}
\label{fig:complexTable}
\end{figure}

\begin{figure}[!htb]
\centering
\resizebox{\linewidth}{!}{
    \includegraphics{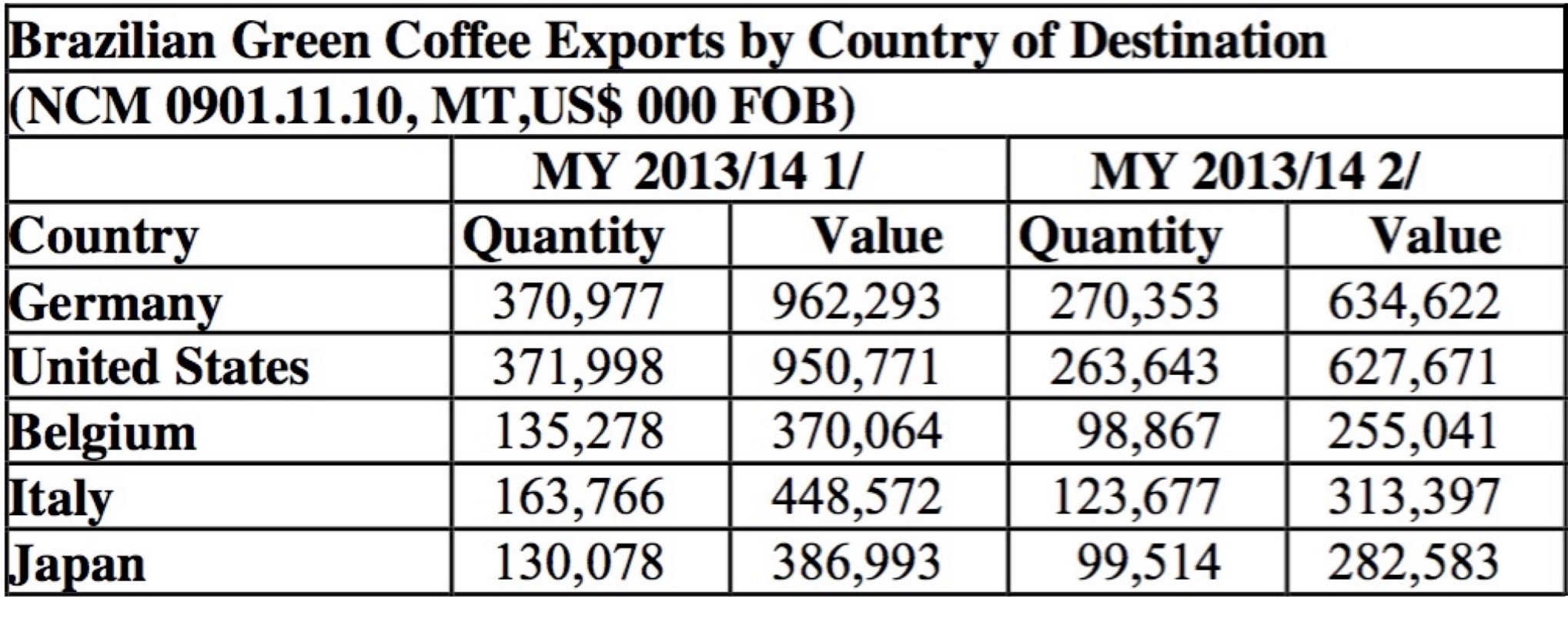}
    }
\caption{Compact table layouts that group values around keys located vertically and horizontally: Coffee quantity/value by year and country.}\label{fig:coffee}
\end{figure}

For this work, we make the following observations: \begin{enumerate*}[label=\textit{(\alph*)}]
\item Many tables that seem complex are in fact compact representations of much simpler and larger tables. 
\item Compacting a simple (flat) but large table into a compact print-layout can be done by grouping repeated values together and using a combination of matrix layouts, spanning cells and nested horizontal or vertical headers for these grouped values. 
For example, Figure~\ref{fig:coffee} shows a matrix layout that groups data by year and country and uses spanning cells to group quantity and value by year. 
In Table~\ref{fig:complexTable}, we see a grouping by internal and external thread fastener arranged as a matrix.
\item Some of these compact tables also include \emph{names} for the values, which sometimes do not have an obvious correspondence to the values in the table.
\item The layout of compact tables is fairly regular. Within same-domain corpora, authors tend to compact their reference data in the same way. Moreover, some of these forms of compacting tables can be seen across corpora of different domains.
\item When people access compact tables, they unconsciously identify the \emph{keys} for the table. However, identifying keys in compact layouts automatically is complicated due to the common practice of introducing column/row names in compact tables to provide hints of their semantics. 
\end{enumerate*}

With these observations in mind, our first objectives are 1) to be able to detect that a table has a compact layout and 2) to \emph{expand} the compact table. Doing so facilitates the analysis of the table using traditional database analysis of the tables. Identifying keys and other dependencies, which comprise key steps in enabling query answering, are described in Section~\ref{sec:queryanswering}
We now describe both processes, starting with table expansion, to lay down some of the notions involved in detecting compact tables which is described in the next subsection.

\subsection{Table Expansion (Flattening)}

The objective of this step is to transform a compact layout (matrix, nested headers, etc.), into an \emph{expanded} representation that is closer to the relation (in the relational theory sense of the word) that the compact table represents. 
The expansion algorithm is based on the assumption that every compact layout can be seen as a matrix layout with three main areas, i.e,. a pair of top and left areas that group values, and a plain data area, which holds the non-grouping values of table.

\begin{figure*}[h!]
    \centering
    \resizebox{0.8\linewidth}{!}{
    \includegraphics{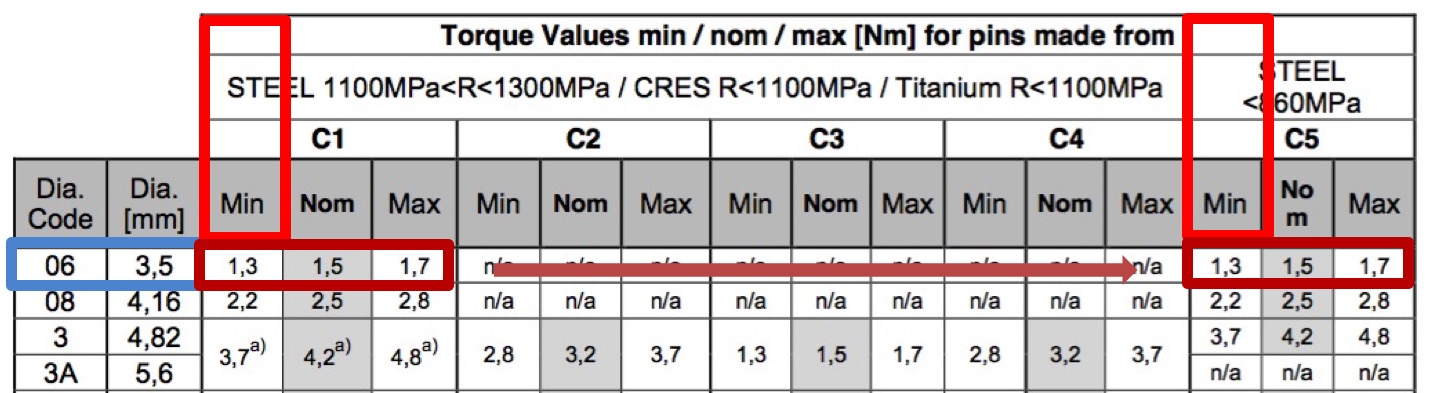}
    }
    \caption{Expansion algorithm example. Horizontal, Vertical and Pivot windows shown in blue, red and dark red, respectively.}
    \label{fig:flattening}
\end{figure*}

Intuitively, the algorithm slides a window called \emph{pivot window} over the main \emph{plain area} of the table. At each step, we produce an \emph{expanded row} by merging the values in the \emph{pivot window} with the values in the \emph{horizontal-axis window} (the first grouping set of values) and the \emph{vertical-axis window} (the second grouping set of values). The pivot window slides one step at a time, together with the vertical-axis window. At each step, a new expanded row is generated. When the end of the current compact row has been reached, the horizontal location of the pivot and vertical windows is reset, and the vertical location of the pivot window goes down by one step. The process is repeated until all of the table has been scanned. A table expansion example is provided in Figure~\ref{fig:flattening}. Note that non-matrix, horizontal tables are just a case of the more general matrix-layout in which there is no horizontal-axis window. 
\begin{figure*}[h!]
    \centering
    \small
            \begin{tabular}{|c|c|c|c|c|c|c|c|c|}
            \hline
            \textbf{column 1} & \textbf{column 2} & \textbf{column 3} & \textbf{column 4} & \textbf{column 5} & \textbf{column 6} & \textbf{column 7} & \textbf{column 8} &  \textbf{column 9} \\
            \hline
            06 & 3,5 & Torq... & STEE... & C1 & Min & 1,3 & 1,5 & 1,7  \\
            06 & 3,5 & Torq... & STEE... & C2 & Min & n/a & n/a & n/a  \\
            06 & 3,5 & Torq... & STEE... & C3 & Min & n/a & n/a & n/a  \\
            06 & 3,5 & Torq... & STEE... & C4 & Min & n/a & n/a & n/a  \\
            06 & 3,5 & Torq... & STEE... & C5 & Min & 1,3 & 1,5 & 1,7  \\
            08 & 4,16 & Torq... & STEE... & C1 & Min & 2,2 & 2,5 & 2,8  \\
            \vdots & \vdots & \vdots &\vdots & \vdots & \vdots &\vdots & \vdots & \vdots \\
             \hline
        \end{tabular}
        \caption{Expansion of table in Figure~\ref{fig:flattening}}
        \label{tab:my_label}
\end{figure*}

Note that expanding in this way, we may generate row values which are not actual values. For example, in Figure~\ref{fig:flattening} values of columns 3 and 6 are actually names for some of the values of the original table. However (as we will see in Section~\ref{sec:queryanswering}
, this \emph{noise} is not problematic during query answering. 
While the name of the expanded columns is not critical for our query answering approach, we try to keep the corresponding names when possible. For hints on where to find the name of columns as well as the parameters of the expansion algorithm (i.e. window locations and sizes), we use the notion of \emph{table families}, described in the following.

 Also note that the size of the pivot window defines the location each sliding step of such window. Parameters such as location of the horizontal and vertical windows are configurable too, e.g., the former may be on the left or right side of the matrix, and the latter may be at the top or bottom of the matrix. 

Note that the flat rows may contain some fields not actual values for the relation described by the original table. For example, each flat row contains a \emph{Min} value. While this is strictly an error (i.e., \emph{Min} is column name for the first column of each pivot window.

\subsection{Table Families and Family Classification:}

As previously mentioned, a key to table expansion is understanding the layout of the table. In particular, understanding which areas of the table are grouping values for expansion, which ones are plain values, etc. Obtaining this information, however, is a challenging task. As we previously mentioned, table layouts in documents can vary wildly. Tables in documents may contain nested tables, images, etc. This complexity is the reason why so many domain-specific table-mapping approaches exist. While a general approach to understand exactly the table layouts seems out of reach at the moment, we opted for an in-between, practical solution which we now describe.

We noted before that corpora on the same domain tend to organize tables in similar ways. While the tables may not have exactly the same form, similar patterns appear. For example, the table in Figure~\ref{fig:complexTable} is a table from our patterns corpus. These tables contain combinations of values for nuts and bolts (external and internal fasteners). While these tables are individually different, they fall in the same pattern. All are matrix tables, with 4-6 header rows that we can consider \emph{horizontal header} rows and 4-6 \emph{vertical header} columns, and where the Nut/Bolt IDs determine the combination values (i.e., C codes), as well as the rest of the features of the nut and bolt. Other tables of the same type may vary in the number of horizontal and vertical header elements, but they all determine one single C code. We call this grouping of tables a \emph{table family}.

Our approach to determine expansion values involves 1) identifying general table families that are of interest to the user, 2) setting the expansion parameters for each family and 3) providing a simple mechanism to identify a table family. Steps 1 and 2 are done on a \emph{per corpus} basis, although, it is possible to define corpus independent families (more on this in Section~\ref{sec:conclusions}). For step 3, we propose a machine learning approach that doesn't require manual annotation of tables, i.e., one of the most cumbersome tasks in traditional table-mapping approaches.

To classify tables into families, we leverage a supervised approach to train a classifier that can help identify the table families. The features we use are domain independent, e.g.,  \textit{number of row headers}, \textit{number of column headers}, \textit{number of repeated column header groups}, \textit{header contains spanning headers} and \textit{number of empty cells on upper left corner of the headers}. 

 While reading a document humans can distinguish the table headers from the data part using two heuristics. Headers typically contain descriptive words as opposed to the data part and therefore they can semantically understand the header content. The second method is to identify the patterns in the table without considering the semantic aspect. For instance header rows typically contain long sequences of alphabetic characters whereas the rest of the columns contain numeric values with similar numbers of digits. Computers can leverage these two approaches to identify the headers. In public domains such as news articles, the semantic understanding approach can be used effectively. However we realized that in complex domains the table headers contain values that are not easily mapped to entity names. Therefore semantic identification methods are not effective on specialized domains. Based on this observation we decided to implement a pattern based approach to identify the headers. 

For the purposes of identifying the headers areas, we first apply a masking methodology to convert the cell values into masked form. In this masked form continuous sets of digits are denoted as `D', continuous sets of alphabetic characters are represented with the character `A' and continuous sets of non alphanumeric chars are represented with `N'. For instance a header string ``Eng FNU-52X'' is converted into a masked form of ``ANDA''. Once the masking is applied on each column based on these rules, distinguishing patterns appear in each column. Typically in at least one column we observe this kind of pattern changes from the header part to the data part. In the data part of each column we check if there are some repeated patterns. If there is such a pattern we conclude that the data part of the table starts at that particular row, and we are able to determine the features mentioned above. In total we used nine table features to classify table families. 

In the experiments with our industrial partner, we had 7 table families. For example, the table in Figure~\ref{fig:complexTable} belongs to the family identified as \emph{matrix-5-by-5-by-1}, that is., a family of matrix tables with a pivot window of width 1 located in cell (5,5). We used about 130 tables from our corpus for the training purposes. 85\% of them were used for training the models and 15\% were used for evaluation. From the Scikit library we used the following classification algorithms to train the models: ``Logistic Regression'', ``Linear Discriminant Analysis'', ``KNeighbors Classifier'', ``Decision Tree Classifier'', ``Gaussian NB'', ``SVC'', ``MLP Classifier'', ``Random Forest Classifier''. 
As a result of k-fold cross validation results, we observed that ``Linear Discriminant Analysis'', ``Decision Tree Classifier'', ``Gaussian NB'' and ``Random Forest Classifier'' classifiers perform quite well on the given training data set, very close to 1. Among these algorithms we decided to go with a naive Bayes classifier since it performed slightly better than the other three classifiers.

\section{Schema description}\label{sec:aboxes_description}

Once the tables are extracted and expanded, we merge data and meta-data for all tables, into a single graph schema that enables structured query answering and indexing of all rows. The use of this schema for query answering is described in the next section. Now we proceed to describe the schema.

\begin{figure*}[h]
\centering
\scalebox{1}{
\begin{tikzpicture}
\node (t)[draw,shape=ellipse,fill=green,minimum width=3cm,minimum height=0.7cm] at (2, 3) {table1};
\node (r)[draw,shape=ellipse,fill=green!60,minimum width=3cm,minimum height=0.7cm] at (4, 0.6) {table1-row1};
\node (a)[draw,shape=ellipse,fill=green!60,minimum width=3cm,minimum height=0.7cm] at (0, 0.6) {table1-column1};
\node (v1)[draw,fill=green!20,anchor=west,minimum width=2.7cm,minimum height=0.6cm] at (8, 3) {cell\_value(1,1)};
\node (v2)[draw,fill=green!20,anchor=west,minimum width=2.7cm,minimum height=0.6cm] at (8, 2) {cell\_value(1,2)};
\node (v3)[] at (9.4, 1.4) {$\vdots$};
\node (v4)[draw,fill=green!20,anchor=west,minimum width=2.7cm,minimum height=0.6cm] at (8, 0.6) {cell\_value(1,N)};
\draw [->]  (r) -- (v1.west);
\draw [->]  (r) -- (v2.west);
\draw [->]  (r) -- (v4.west);
\draw [->]  (t) -- (r);
\draw [->]  (t) -- (a);
\node[rotate=0,scale=0.8,color=blue] at (2,3.5){Table};
\node[rotate=0,scale=0.8,color=blue] at (0,0.15){Attribute};
\node[rotate=0,scale=0.8,color=blue] at (4,0.15){Row};
\node[rotate=0,scale=0.8,color=blue] at (9.4,3.48){xsd:string};
\node[rotate=0,scale=0.8,color=blue] at (9.4,2.48){xsd:string};
\node[rotate=0,scale=0.8,color=blue] at (9.4,0.12){xsd:string};
\node[rotate=31,scale=0.8] at (6.2,2.1){table1-column1};
\node[rotate=19,scale=0.8] at (6.6,1.67){table1-column2};
\node[rotate=0,scale=0.8] at (6.8,0.75){table1-columnN};
\node[rotate=-50,scale=0.8] at (3.15,1.85){hasRow};
\node[rotate=50,scale=0.8] at (0.9,1.9){hasAttribute};

\end{tikzpicture}
}
\caption{ABoxes generated via our system for one row of a flattened table. (Note that table columns are instances of the class Attribute and properties at the same time)}
\label{fig:aBoxes}
\end{figure*}
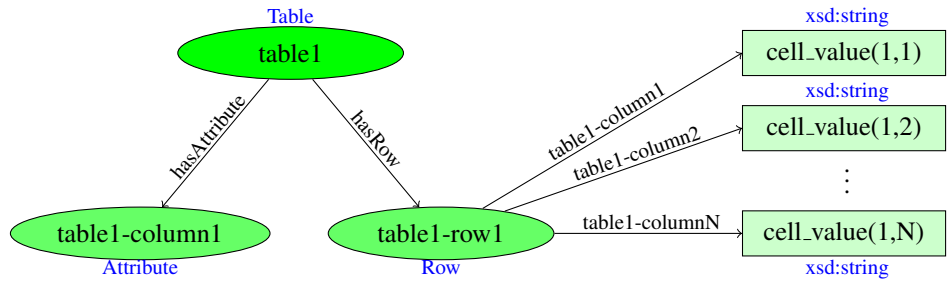

We divide the elements of this schema in two categories: the first one about the components that are given in the expanded tables (e. g. rows and columns) and the second one about the elements that are retrieved afterwards (e.g. dependencies and keys). An overview of the RDF types and properties we used is presented in the the following two sections and summarized in Figure~\ref{fig:tBoxes}. In Figure~\ref{fig:aBoxes} we present instead an example of ABoxes generated using our system, for one row of a expanded table.

\begin{figure*}[h]
\centering
\resizebox{\linewidth}{!}{
\begin{tikzpicture}
\node (a)[shape=ellipse,draw,fill=blue!30] at (8, 6) {Document};
\node (s)[shape=ellipse,draw,fill=blue!30] at (8, 4) {Table};
\node (d)[shape=ellipse,draw,fill=blue!30,minimum width=4cm] at (8, -0.25) {Attribute$^b$};
\node (f)[shape=ellipse,draw,fill=blue!30] at (10, 1) {Row};
\node (g)[draw,fill=black!20,anchor=west] at (12, 1.33) {xsd:string (caption)}; 
\node (h)[draw,fill=black!20,anchor=west] at (12, 2.66) {xsd:integer (pageBegin)}; 
\node (j)[draw,fill=black!20,anchor=west] at (12, 3.99) {xsd:integer (pageEnd)}; 
\node (k)[draw,fill=black!20,anchor=west] at (12, 5.33) {xsd:string (ID)}; 
\node (l)[draw,fill=black!20,anchor=west] at (12, 6.66) {xsd:string (label)}; 
\node (p)[draw,fill=black!20] at (10, -1.5) {xsd:string (cellValue)}; 
\node (q)[shape=ellipse,draw,fill=blue!30] at (1.5, 6) {Dependency};
\node (w)[shape=ellipse,draw,fill=blue!30] at (4, 4) {Key};
\node (e)[shape=ellipse,draw,fill=blue!30] at (4,2) {CompositeKey};
\node (r)[shape=ellipse,draw,fill=blue!30] at (0.2,4) {InclusionDependency};
\node (t)[shape=ellipse,draw,fill=blue!30] at (0.2,2) {AttributePair};
\draw [->]  (w) -- (q);
\node[rotate=-39,scale=0.8] at (3,5){subClassOf};
\draw [->] (e) -- (w);
\node[rotate=57,scale=0.8] at (0.68,5){subClassOf};
\draw [->]  (r) -- (q);
\node[rotate=0,scale=0.8] at (4,3){subClassOf};
\draw [->]  (q) -- (s);
\node[rotate=-17,scale=0.8] at (4.85,5.1){hasDependency};
\draw [->]  (a) -- (s);
\node[rotate=0,scale=0.8] at (8,5){hasTable};
\draw [->]  (s) -- (d);
\node[rotate=-90,scale=0.8] at (8.2,2){hasAttribute};
\draw [->] (w) to [bend left](d.north);
\node[rotate=0,scale=0.8] at (6.5,3){hasComponent};
\draw [->] (r) to (t);
\node[rotate=0,scale=0.8] at (0.2,3){hasAttributePair};
\draw [->] (s) to (f);
\node[rotate=-56,scale=0.8] at (9.2,2.5){hasRow};
\draw [->] (s) to (g.west);
\node[rotate=-34,scale=0.8] at (10.5,2.5){hasCaption};
\draw [->] (s) to (h.west);
\node[rotate=-18,scale=0.8] at (10.5,3.3){pageNumBegin};
\draw [->] (s) to (j.west);
\node[rotate=0,scale=0.8] at (10.5,4.1){pageNumEnd};
\draw [->] (s) to (k.west);
\node[rotate=17,scale=0.8] at (10.5,5){hasID};
\draw [->] (s) to (l.west);
\node[rotate=32,scale=0.8] at (10.3,5.7){rdfs:label};
\draw [->] (e) to (d);
\node[rotate=-29,scale=0.8] at (6.3,0.8){hasComponent$^a$};
\draw [->] (t) to [bend right](d.south);
\node[rotate=-21,scale=0.8] at (3.8,0.7){secondComponent};
\draw [->] (t) to (d.west);
\node[rotate=-19,scale=0.8] at (3.8,-0.4){firstComponent};
\draw [->] (f) to (p);
\node[rotate=-90,scale=0.8] at (10.2,-0.3){Attribute$^b$};
\end{tikzpicture}
}
\caption{T-boxes of our system. {$^a$CompositeKey has at least two instances of hasComponent, while Key has only one}. {$^b$Instances of Attribute are both properties and instances of classes (common practice in the RDF community).
}}
\label{fig:tBoxes}
\end{figure*}
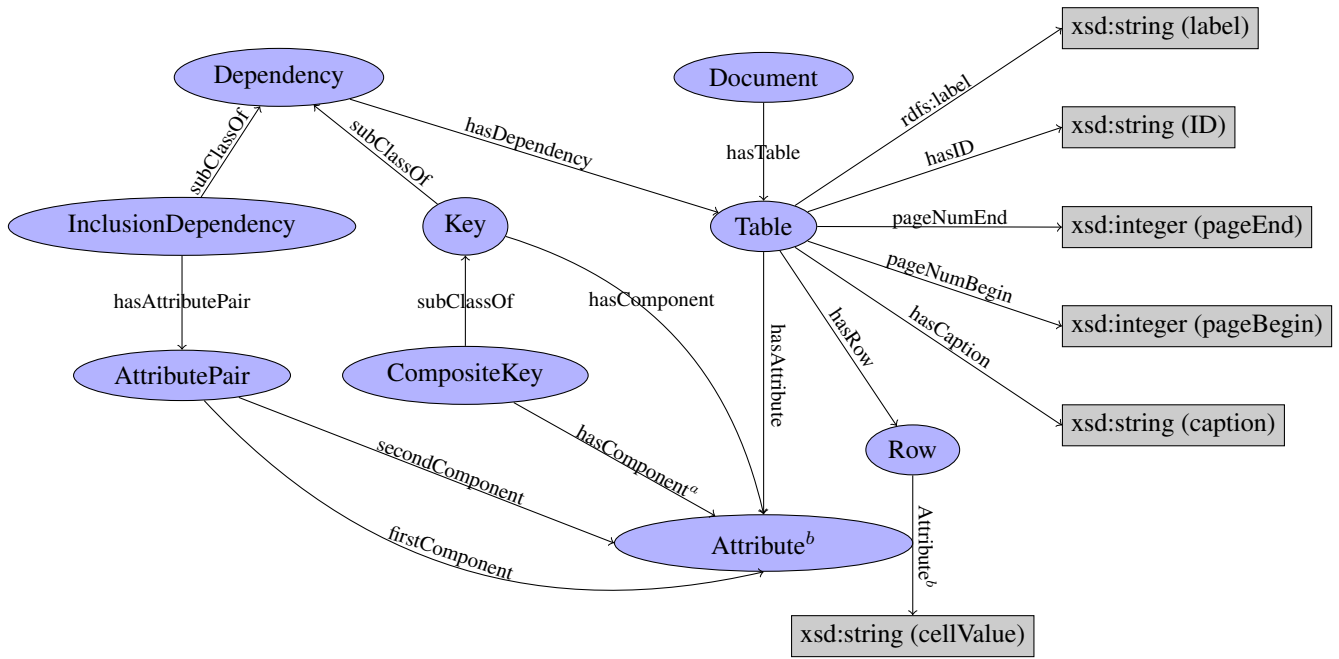

\subsection{Schema for table data}
The following are the classes (i.e., types) and properties for the objects that represent tables and their rows in our graphs:
\begin{itemize}
\item{\bf Document} identifies objects that are {\it documents}, i.e., PDF source files that were previously ingested. They may be connected to one or more Table objects.
\item{\bf Table} identifies objects that are {\it expanded tables}, i.e., relational structures with attributes, rows and dependencies. Table objects are always associated with a Document object, one or more Attribute objects, one or more Dependency objects, and one or more Row objects.
using the properties hasTable, hasAttribute, hasDependency, and hasRow. Tables may include a caption, an ID, and page numbers using the properties hasCaption, hasID, pageNumBegin, pageNumEnd. 
\item{\bf Attribute} identifies object that are {\it attributes}, e.g. columns of tables. Attribute objects may or may not have a label (usually obtained from the header of the attribute); 
\item{\bf Row} identifies objects that are {\it rows} of a table and which will be associated to literals containing the values for each cell of the row in the expanded table.
\end{itemize}

The following are the properties we use to describe documents, tables and rows in our schema:
\begin{itemize}
\item{\bf hasTable} connects a Document with zero or more Table objects;
\item{\bf hasAttribute} connects a Table object with one or more Attribute objects;
\item{\bf hasRow} connects a Table object with one or more Row objects;
\item{\bf hasCaption} connects a Table object with a xsd:string value;
\item{\bf hasID} connects a Table object with a xsd:string value;
\item{\bf pageNumBegin} connects a Table object with a xsd:integer value;
\item{\bf pageNumEnd} connects a Table object with a xsd:integer value; and
\item{\bf hasParentLabel} connects a Attribute object with a xsd:string value.
\end{itemize}

\subsection{Schema for dependencies}
\label{sec:aboxes_dependencies}
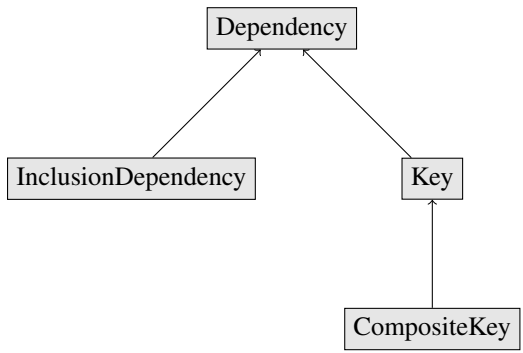
\begin{figure}[h]
\begin{center}
\begin{tikzpicture}[scale=1]
\tikzstyle{every node}=[draw,fill=gray!20];
\node (d) at (2, 4) {Dependency};
\node (k) at (4, 2) {Key};
\node (ck) at (4,0) {CompositeKey};
\node (id) at (0,2) {InclusionDependency};
\draw [->]  (id) -- (d);
\draw [->] (ck) -- (k);
\draw [->]  (k) -- (d);
\end{tikzpicture}
\caption{Hierarchy between classes: Dependency, Key, CompositeKey and InclusionDependency}
\label{fig:dependency_hierarchy}
\end{center}
\end{figure}

For each table, we extract the set of all minimal keys using existing techniques~\cite{Ramakrishnan:2002:DMS:560733}. For all tables within a document, we are interested in inclusion dependencies to determine the relations between pairs of tables an enable multi-table queries (see Section~\ref{sec:queryanswering}). 
The following are the classes that we use to define dependencies over a table object:
\begin{itemize}
\item{\bf Dependency} identifies an object as a {\it dependency};
\item{\bf Key} and {\bf CompositeKey} are placeholders for a set of columns (attributes) that functionally determine the values of the rows in a table (Key has exactly one Attribute and CompositeKey has more than one Attribute);
\item{\bf InclusionDependency} is a placeholder for a set of {\it pairs of columns}. Each list contains a column in the current table and a column of another table (if the columns for the second table are part of a Key, then the inclusion dependency is also a foreign key); and
\item{\bf  AttributePair} is a placeholder for a pair of Attribute objects, used to form foreign keys.
\end{itemize}
In the ABoxes we assume the rdfs:subClassOff hierarchy, regarding dependencies and keys, described in Figure~\ref{fig:dependency_hierarchy}

We use the following properties we used to describe dependencies in our schema:
\begin{itemize}
\item{\bf hasDependency} connects a Table object with one or more Dependency objects;
\item{\bf hasComponent} connects a Key object or CompositeKey object with Attribute objects;
\item{\bf hasAttributePair} connects a InclusionDependency objects with one or more AttributePair objects; and
\item{\bf firstComponent} and {\bf secondComponent} connect a AttributePair object with a Attribute object.
\end{itemize}

\section{Query Answering}\label{sec:queryanswering}

The kind of queries we envision are queries where the user does not need to know the underlying schema -- she only needs to specify some features of the data that she is looking for, such as values for some of the properties or the name/label of an object. Our techniques find the relevant table or tables and provide a list of rows that best satisfy the query. We provide a simple query language illustrated in Figure~\ref{fig:samplequery}. 

The fields in the input query are: 
\begin{itemize}\item\textbf{table\_caption}, a list of topics used to identify relevant tables based on their captions; 
\item\textbf{projection}, a list of string descriptors of the attribute(s) the user seeks a value for; and \item\textbf{conditions}, a list of (\textit{attr}, \textit{val}) pairs that indicate describe the selection criteria of a result in terms of an \emph{attribute name} and a \emph{value}. 
\end{itemize}
Almost all parameters are optional, except values for conditions. 
\begin{figure}
\centering
\begin{lstlisting}[basicstyle=\ttfamily\scriptsize,frame=single,showstringspaces=false]
{ table_caption:"Torque values",
    projection:["Max"],
    conditions:[{attribute_name:"Dia. Code", value:"3A"}, 
       {value:"C2"}]}
 \end{lstlisting}
\caption{Sample query in our structured format which retrieves answer from Table~\ref{tab:my_label}}\label{fig:samplequery}
\end{figure}
As the user may not know the structure of the original tables, our search algorithm attempts to align query input to values in the tables using exact/fuzzy match and leverages table metadata to further refine the search (Section~\ref{sec:aboxes_description}).

For simple queries that can be answered using a single table we rank the results using the contents and metadata of the table while tolerating under-specified queries (see Section~\ref{sec:single-table-query-answering}). Answering complex queries may require inferring missing inputs and/or joining rows across multiple tables (see Section~\ref{sec:dependencies-query-answering}).

 \subsection{Single-table query answering}\label{sec:single-table-query-answering}
When answering queries we translate the input query into a SPARQL query that searches (available in most triple-store today) for the provided conditions over the rows and meta-data of all expanded tables in our graph. The results are ranked according to a scoring function based on the triple store's built-in \texttt{search} capabilities. The scoring function is constrained so that fuzzy matches are preferred over unmatched fields. The best method of composing a score for each row may depend on the application; we chose a weighted sum over the matching score for each field query.

Our \texttt{WHERE} clause specifies the type constraints on the \texttt{?row} variable along with a \texttt{?table :hasRow ?row} relation. If \textbf{table\_topics} is non-empty, we add an \texttt{OPTIONAL} construct attempting to match each aspect of the input query. For each attribute descriptor in the \textbf{projection} list we create a variable \texttt{?proj\_i} constrained to be an attribute in the table and attempt to match on its label. Finally, \texttt{attr\_i} and \texttt{val\_i} variables are created for each of the \textbf{condition} pairs by an \texttt{OPTIONAL} construct that constrains them to appear in the same row: \texttt{?row ?attr\_i ?val\_i}. The matching score on their labels is computed in the same way as before.

Figure~\ref{fig:querysingletable1} illustrates the translation of the query in Figure~\ref{fig:samplequery} in which a user is attempting to learn the maximum torque value for a screw with diameter code 3A and combination code C2 for some unknown attribute in some table in the corpus. 

\setlength{\belowcaptionskip}{-5pt}
\begin{figure}
\centering
\begin{lstlisting}[basicstyle=\ttfamily\scriptsize,frame=single,showstringspaces=false]
SELECT ?row ?proj_val_1 WHERE {   
   ?row rdf:type :Row .
   ?table :hasRow ?row .
   OPTIONAL { ?table :hasCaption ?cap_1 .
              ?cap_1 search "Torque values" }
   ...
   OPTIONAL { ?table :hasAttribute ?proj_1 .
              ?proj_1 search "Max" .
              ?row ?proj_1 ?proj_val_1 }
   ...
   OPTIONAL { ?row ?attr_1 ?val_1 .
              ?attr_1 search "Dia. Code" .
              ?val_1 search "3A" }
   OPTIONAL { ?row ?attr_2 ?val_2 .
              ?val_2 search "C2" }  }
\end{lstlisting}
\caption{Sample $QuerySingleTable$ output for a query on the table shown in Figure~\ref{fig:flattening}.}
\label{fig:querysingletable1}
\end{figure}

\subsection{Using dependencies for query answering}\label{sec:dependencies-query-answering}
In this section we explore the problem of answering a user's query using dependency information across a set of tables. Our method can be used in several contexts and is particularly useful in scenarios in which we don't have explicit information about the structure of the schema that is underneath the given data and we want to enable answers to queries that require joining multiple tables. 

After the tables have been expanded and indexed, dependencies in the form of keys or inclusion dependencies allow the system to provide schema-less query methods that facilitate access to the data.

The main use case for our approach is for processing queries on documents that include tables (such as PDF corpora), which, by their nature, don't contain fixed schemas; another use case is the exploration of large datasets such as those crawled from the web.

\subsubsection{The method}
Our input is a set of expanded tables, for example the ones we saw in Section~\ref{sec:flattening_familyDetection}
, onto which has been added information about dependencies (see Section~\ref{sec:aboxes_dependencies}
). In this context, dependencies and keys allow us to resolve complex queries on the data. This scenario occurs when the input provided by the user's query is not sufficient to define a single answer and/or the query involves the contents of multiple tables. We use dependencies to complete keys by recovering values that are not directly specified in the input but which can be retrieved using the values that are. We can use foreign dependencies (which define the relationship of a table column with respect to the columns of other tables) to connect data stored across multiple tables. Moreover, in the event where the initial query produces more than one unique solution, our system can suggest the minimal set of parameters for which specifying a value would obtain a unique answer.

This method is applicable to tables and queries that involve answers from single tables or multi-tables (i.e. queries that require joining two or more tables to compute an answer). The algorithm takes as input a collection of A-box triples regarding tables and their information regarding dependencies and keys.
The idea behind our system is to pre-compute (offline) all the possible keys and foreign keys for each table. At run time, given an input query the system uses these keys and dependencies to search for the possible answer(s) to the query.

The system uses the following steps:
1) It checks if there exists at least one table that contains values that match the user query. In this case, the system returns the corresponding row(s) to the user as its answer.
2) If there does not exist any table that contains all the input values, then the system will:
2a) identify the tables of interest and
2b) iterate over these tables, using both foreign dependencies and functional dependencies to try to recover the values of each key (in order) of the current table. This corresponds to the inference step; we use dependencies as logical rules trying to retrieve the key values.
2c) If it is possible to recover at least one key, then we have found a solution and return the corresponding row(s) to the user.
2d) If there is no solution, then all the rows containing the input data are returned to the user (an incomplete solution).

A more detailed description of the used algorithm is presented in Algorithm \ref{alg:query_dependencies}, {\it QueryMultiTable}, which uses $QuerySingleTable$ from Section~\ref{sec:queryanswering} to perform a SPARQL query (possibly on a particular table) for a row given a set of values.

A more detailed description of the used algorithm is presented in Algorithm \ref{alg:query_dependencies} {\it QueryMultiTable}, which uses $QuerySingleTable(table,values\_list)$ to perform a single SPARQL query on a table for a row with values in $values\_list$ and $QuerySingleTable(\_,values\_list)$ perform a single SPARQL query for a row with values in $values\_list$ in any table as follows:
\begin{figure}
\centering
\begin{mdframed}
\begin{lstlisting}[basicstyle=\ttfamily\scriptsize,showstringspaces=false]
SELECT ?row WHERE {   
   ?v1+ search* value1.            
   ?row ?property1 ?v1.
   ?v2+ search* value2.           
   ?row ?property2 ?v2.
   ...
   <table_URI> krt:hasRow ?row. or ?table krt:Row ?row.
   ?row rdf:type krt:Row
}
\end{lstlisting}
\end{mdframed}
\caption{QuerySingleTable *where \texttt{search} is the indexed search function embedded in the third-party RDF store.}\label{fig:querysingletable}
\end{figure}
\begin{algorithm}[h]
\footnotesize
\caption{$QueryMultiTable$}
\label{alg:query_dependencies}
\SetAlgoVlined
\SetArgSty{}
\KwIn{$goal\_labels$, $input\_values$}
$results$= $QuerySingleTable(\_,input\_values)$\;
\uIf{$results \not= null$}{
    \KwRet{$results$}}
\uElse($results == null$){
    $tables$=Search all the tables containing goals labels\;
    Reorder $tables$ following the number of known elements in the table\;
    \For([following the new order]){$table$ in $tables$}{
        $keys$ = Find the primary keys for $table$\;
        \For{key in keys}{   
            $key\_components$ = Retrieve $key$ components\;
            $known\_components$= Identify the known elements of the key\;
            $unknown\_components$= ($key\_components$) $\setminus$ ($known\_components$)\;
            instantiate $retrieved\_values$\;
            \For{$unknown\_component$ in $unknown\_components$}{   
                $retrieved\_values$+= $RetrieveValues(table,$
                   $unknown\_component,$ 
                   $(input\_values \setminus known\_components))$
                   [see Algorithm \ref{alg:query_RetriveSingleValue}]\;
            }
            $results$+=$QuerySingleTable(table,$
               $(input\_values,retrieved\_values))$\;}
    }
    \KwRet{$results$}\;
}
\end{algorithm}
\begin{algorithm}[h]
\footnotesize
\caption{$RetrieveValues$}
\label{alg:query_RetriveSingleValue}
\SetAlgoVlined
\SetArgSty{}
\KwIn{$table$, $unknown\_component$, $values$}
$retrieved\_values$=null\;
$inclusion\_dep$=Find all the inclusion dependencies for the $unknown\_component$\;
\For{($dependency\_table$,$dependency\_column$) in $inclusion\_dep$}{
$result\_row$= $QuerySingleTable(dependency\_table,column,values)$ = rows with the remaining not used input data\;
$retrieved\_value$=projection of $result\_row$ on $dependency\_column$\;
$retrieved\_values+=retrieved\_value$\;
}
\KwRet{$retrieved\_values$}\;
\end{algorithm}
We note that this method is also applicable to object graphs and knowledge graphs, performing a projection of the data to a database (for example if we see each object as a ``row'' and the values for its properties as ``row cells'' or ``row values''). Using our techniques, we may be able to develop more efficient methods for finding ``long range'' dependencies between objects in the graph and leveraging those dependencies to answer queries.

\section{Evaluation}\label{sec:evaluation}

The approach presented here has been developed in the context of a project with a large industrial partner. We used their use-case and data to show the feasibility and benefits of our approach. We now proceed to describe this use-case.

Our industrial partner uses thousands of tables embedded within PDF documents. These tables are used as reference material by technicians in assembly lines and equipment maintenance locations. During their activities, technicians use these tables to look up information for their tasks. Given that these look-ups require either a physical search on a printed PDF document or a plain keyword search over the PDFs, finding the right information often requires minutes. Look-ups of this form are a routine operation, which summed up across all technicians in all locations of our client, amounts to thousands of hours per year for our partner. 
Our objective is to reduce the cost of each of these individual look-ups by offering a structured query language that would allow users to query these documents similarly to the way they would query regular databases, but with very little knowledge of the schema. 

Our partner focused on two tasks, \textit{single-table retrieval}, that is, retrieving information that lives in a single table; and \textit{multi-table retrieval}, that is, retrieving information that requires consulting multiple tables. 
An example of single-table query is, \emph{given an internal and external thread fastener such as ITF14 and ETF2, retrieve the corresponding combination code}. This kind of query returns answers from tables similar to the one shown in Figure~\ref{fig:complexTable}. There are hundreds of these tables within multiple documents. 
A second example of single-table queries is \emph{given a diameter code of 8, and a combination value of C2, retrieve the minimum, nominal and maximum torque tolerance values}. This kind of query has answers from tables similar to the one shown in Figure~\ref{fig:flattening}. Similarly, there are hundreds of these tables, in multiple documents.
Finally, multi-table queries are of a form which is a combination of the previous queries, for example, \emph{given the codes for an internal and external thread fastener, ITF14 and ETF2, and a diameter code of 08, retrieve the minimum, nominal and maximum torque tolerance values}. This kind of query requires the system to discover the need for joining two tables of the kind presented in Figures~\ref{fig:complexTable}~and~\ref{fig:flattening}. 
\begin{figure}
\centering
\begin{mdframed}
\centering
\minipage{0.5\textwidth}
\centering
\begin{lstlisting}[basicstyle=\ttfamily\scriptsize,showstringspaces=false]
{conditions:[
  {value:"ITF14"},
  {value:"ITF2"}]}
{conditions:[
  {value:"8"},
  {value:"C2"}]}
\end{lstlisting}
\endminipage
\minipage{0.5\textwidth}
\centering
\begin{lstlisting}[basicstyle=\ttfamily\scriptsize,showstringspaces=false]
{targets:[
  "min","nom","max"],
  conditions:[
  {value:"8"},
  {value:"ITF14"},
  {value:"ITF2"}]}
\end{lstlisting}
\endminipage
\end{mdframed}
\caption{Sample single and multi-table queries}\label{fig:querysingletable3}
\end{figure}

\section{Related Work}

The problem of query answering over tables can be partitioned into two dimensions; first extracting and understanding the semantics of the table’s data, and second, offering flexible ways to query the data without having the knowledge of the table schema. Below we provide an overview of past work on these two angles.

\subsection{Extracting and understanding the semantics of the table’s data.}
There are many studies in the literature as well as products targeting the problem of table extraction from digital documents. However, in our experiments, we observed that while these methodologies are quite effective on general domain articles they perform quite poorly when it comes to extraction of tables from complex domain specific documents.
\citet{Liu:2007:TAT} and \citet{Liu:2006:AET} introduce a new tool called TableSeer for automatic metadata extraction from digital documents. TableSeer crawls digital libraries, detects tables from documents, extracts metadata, indexes and ranks tables, and provides a user-friendly search interface. TableSeer extracts the data from tables using predetermined heuristics which limit the information extraction capabilities for domain specific complex tables. Our machine learning based learning approach enables custom training for specific document corpora and provides much higher precision when it comes to extraction of table content. 
\citet{Lipinski:2013} focus on the problem of metadata extraction from scientific articles and evaluate the performance of existing tools. They mention that some of these tools use heuristics while others leverage machine learning algorithms such as support vector machines (SVM), hidden Markov models (HMM), or conditional random fields (CRF). However, none of these approaches leverage these ML methods for table classification and flattening these table structures. 
\citet{Ramakrishnan2012} propose a layout aware text extraction method from PDF documents which doesn't have a particular focus on table content extraction. 
\citet{Pinto:2003:TEU} propose a method for identifying header separators and columns separated with white spaces. Our work is different from this work as we particularly focus on extraction of hierarchical headers and flattening them using a pivot-based approach. 
There have been several papers on extracting tabular data from web tables.
Google has done a considerable amount of work in this area~\cite{balakrishnan2015applying}. WebTables is a system to extract and leverage the relational information embedded in HTML tables on the web~\cite{cafarella2008webtables}. The paper describes how to use the attribute correlation statistics database which is a set of statistics about schemas in the corpus. 
\citet{Tengli:2004:LTE} propose algorithms to extract information from HTML-based web tables. The algorithms leverage the HTML tags to identify spanning headers and rows. But the proposed approaches do not leverage any machine learning based table classification for table flattening. 
\citet{Cafarella:2009:WES} propose three extraction systems that can be operated on the entire Web. The TextRunner system focuses on raw natural language text, the WebTables system focuses on HTML-embedded tables, and the deep-web surfacing system focuses on ``hidden'' databases.
In the work by ~\citet{pimplikar2012answering}, the problem of extracting relevant columns from several matching web tables and consolidating
them into a single structured table is addressed. The system returns a  multi-column table in response to a query specifying a
set of column keywords.
\citet{Wang12} focus on the problem of understanding the concepts in HTML-based web tables and connecting them to larger taxonomies. Our work is different from~\citet{Wang12} in multiple aspects. First and foremost we particularly focus on domain specific complex tables where extraction of knowledge is not trivial. Once the information is extracted, we also find dependencies among tables in the same corpus and support multi-table queries without having schema information about the tables. In the work by ~\citet{NishidaSHM17}, a table classification method is proposed using deep neural network techniques. However the classification method is not used for flattening complex tables which makes it different from our focus. 
\citet{Ramel:2003:DER} focus on table identification and extraction where cell lines are not clear. Their work is orthogonal to what we are aiming in our work. Extracting information from tables for answering questions is addressed in~\citet{wei2006table}. First, tables are extracted and then specific cells of tables related to answering the questions are retrieved. Information is extracted using conditional random fields.
In the works of ~\citet{Burdick11}, \citet{BalakrishnanCHHKLPPPRSSTVY10}, \citet{BurdickEKLPRW14} an information extraction system has been proposed. 

\subsection{Querying tables without schema information.}
A schemaless query support for graph databases is proposed by~\citet{Yang:2014:SSG}. Our work is different since we discover the dependencies to support multi-table join based queries on the produced graph. Similar to~\citet{Yang:2014:SSG} a schema-free query methodology is proposed in \citet{Zheng:2016:SSS}. In~\citet{Zheng:2017:NLQ}, a natural language question answering system is proposed. In~\citet{Yuan:2016:EPM}, a pattern search mechanism is studied in knowledge graphs. None of these studies focuses on discovering the relationship among tables to leverage that information for query answering. \citet{DBLP:conf/semweb/HassanzadehWRS15} focus on the problem of entity linking between web tables and existing public knowledge graphs like DBpedia Ontology, Schema.org, YAGO, Wikidata, and Freebase. Entity linking with external ontologies is orthogonal to our work and will improve our system as well. 
\citet{Embley:2005:AED} study instead the problem of querying HTML tables without the knowledge of their schemas. The authors try to map existing tables into a targeted schema which makes it different from our work since we do not make an assumption that we have schema information known in advance. 
\citet{Chen:2000:MTL} is similar to \citet{Embley:2005:AED} since the work involves extracting HTML tables but without leveraging ML based extraction methodologies. 

\section{Conclusions and Future Work}\label{sec:conclusions}

In this paper we introduce new methods for users to query data stored in tables within documents. Our methods require minimal effort and are able to handle tables that are of considerable complexity. The techniques presented here exploit the flexibility of the RDF data model, the features of the SPARQL query language and dependency theory. The methods described here have been tested and validated in the context of collaboration with a large industrial partner. 

The first major directions for follow ups of this work are those in the area of data integration. In particular, we foresee that the use of ontology-mapping techniques to integrate the schemas of the flat tables we produce. In the area of table extraction and transformation, future work includes the use of deep learning to further generalize the methods for family detection and/or discovery of the parameters for the flattening algorithms. In the context of schemaless query answering with dependencies, we believe that the use of \emph{probabilistic dependencies} would allow our methods to deal with noisy data, e.g., tables in which there are errors in the extraction process.


\bibliographystyle{aaai}
\bibliography{biblio,base}

\end{document}